\documentclass[aps,prb,twocolumn,floatfix]{revtex4}
\usepackage{amsmath}
\usepackage{natbib}
\usepackage{cases}
\usepackage[pdftex]{graphicx}
\usepackage[normalem]{ulem}

\usepackage[usenames]{color}

\begin{document}

\date{\today}

\author{Seyyed Jabbar Mousavi$^{1,2,a}$, Megan F. Biggs$^3$, Jeremy A. Johnson$^3$, Peter Hamm$^1$ and Andrey Shalit$^{1,a}$}

\affiliation{$^1$Department of Chemistry, University of Zurich, Winterthurerstrasse 190, CH-8057 Zurich, Switzerland}
\affiliation{$^2$Institute of Applied Physics, University of Bern, Sidlerstrasse 5, 3012 Bern, Switzerland}
\affiliation{$^3$Department of Chemistry and Biochemistry, Brigham Young University, Provo, Utah 84602, USA}
\affiliation{$^a$Authors to whom correspondence should be addressed: jabbar.mousavi@unibe.ch, andrey.shalit@chem.uzh.ch}

\title {${\chi}^{(2)}$-Induced Artifact Overwhelming the Third-Order Signal in 2D Raman-THz Spectroscopy of Non-Centrosymmetric Materials}

\begin{abstract}

Through comprehensive data analysis, we demonstrate that a ${\chi}^{(2)}$-induced artifact, arising from imperfect balancing in the conventional electro-optic sampling (EOS) detection scheme, contributes significantly to the measured signal in 2D Raman-THz spectroscopy of non-centrosymmetric materials. The artifact is a product of two 1D responses, overwhelming the desired 2D response. We confirm that by analyzing the 2D Raman-THz response of a x-cut beta barium borate (BBO) crystal. We furthermore show that this artifact can be effectively suppressed by implementing a special detection scheme. We successfully isolate the desired third-order 2D Raman-THz response, revealing a distinct cross-peak feature, whose frequency position suggests the coupling between two crystal phonons. 
 
\end{abstract}

\maketitle

\section{Introduction}

Two-dimensional (2D) THz spectroscopy intends to measure the third-order nonlinear signal, giving useful insights into the dynamics and coupling of low-frequency modes in the THz frequency range in condensed-phase molecular systems.\cite{Kuehn_2009,Woerner_2013,Elsaesser_book,Houver_2019,Woerner_2021} Recently, the application of this technique was further expanded to explore the correlation and dynamics of rotational degrees of freedom in gas-phase molecules.\cite{Fleischer_2012,Nelson_2016,Lu_2019} In two-pulse 2D THz experiment on non-centrosymmetric materials, a ${\chi}^{(2)}$ nonlinear susceptibility can also contribute to the signal.\cite{Reimann_2021} Various second-order contributions, with characteristic spectral features, originating from second harmonic generation (SHG), and optical rectification (OR) processes have been previously reported in 2D THz spectra of non-centrosymmetric materials.\cite{Woerner_2013,Elsaesser_book,Somma_2014, Lu_2017} 

For example, Elsaesser and co-workers studied intersubband transitions in GaAs/AlGaAs semiconductor quantum wells using two-pulse 2D THz spectroscopy.\cite{Kuehn_2011,Kuehn_2011_2} In addition to the third-order pump-probe and photon echo signals, they also observed second-order contributions to the 2D THz spectrum. The observed ${\chi}^{(2)}$ features were assigned to difference-frequency mixing in the bulk GaAs substrate. Nelson and co-workers investigated nonlinear responses of magnons and their correlations in an antiferromagnetic crystal, YFeO$_3$. They also observed that the second-order signals, including SHG and THz rectification (TR) peaks, originated from sum- and difference-frequency mixing of the magnons, respectively, contribute to the measured 2D THz spectra.\cite{Lu_2017} In a very recent 2D THz experiment on ZnTe, it has been shown that second-order contributions, such as rectified nonlinearities and second harmonic signals arising from the electro-optic sampling (EOS) in the detection crystal, known as artifacts of the detection scheme, can also contaminate 2D spectra collected in a collinear geometry.\cite{Liu_2024}  Although a double-chopping scheme, modulating two incident THz fields at 1/2 and 1/3 of the laser repetition rate with lock-in detection of the nonlinear signal at 1/6, was used in this work, strong second-order artifacts due to the EOS were still observed.

Hybrid spectroscopies with two complementary pulse sequences of 2D Raman-THz-THz (2D RTT) \cite{Savolainen_2013,Shalit_2017,Ciardi_2019,Shalit_2021,Mousavi_2022} and 2D THz-THz-Raman (2D TTR) \cite{Blake_2016,Blake_2017,Blake_2020,Knighton_2019,Lin_2022} are alternative approaches for performing 2D experiments in the THz frequency range. In this work, we present a new observation on how a ${\chi}^{(2)}$-induced artifact can dominate the third-order signal in 2D Raman-THz experiments involving non-centrosymmetric samples. The artifact is effectively the product of two linear (i.e., 1D) responses arising from imperfect balancing in the conventional EOS detection scheme. It can overwhelm the desired nonlinear 2D signal, even when a double-chopping scheme\cite{Mickan2002, Strait2009, Iwaszczuk2009} is used, potentially leading to misleading interpretations. Since this artifact is inherent to EOS, it may not appear in measurements using an air-biased coherent detection scheme\cite{Karpowicz2009}, such as in transient THz spectroscopy that employed a double modulation technique with a specific modulation pattern.\cite{DAngelo2016} 

We will first present our measured 2D Raman-THz data for an x-cut BBO nonlinear crystal. By analyzing the measured data, we confirm that the observed features originated from the second-order contribution rather than the third-order response of the sample. We also introduce the origin of this contribution and explain why a double-chopping scheme is insufficient to isolate the third-order signal in such cases. Finally, with the goal of recovering the real 2D Raman-THz response of the BBO crystal, we measured the x-cut BBO signal using a special detection scheme based on the work of Johnson \textit{et al.}\cite{Johnson_2014} and Brunner \textit{et al.}\cite{Brunner_2014}. Our data provide experimental evidence for phonon-phonon coupling in the BBO crystal.

\section{Methods}

\begin{figure}[t]
	\centering
	\begin{center}
		\includegraphics[width=0.45\textwidth]{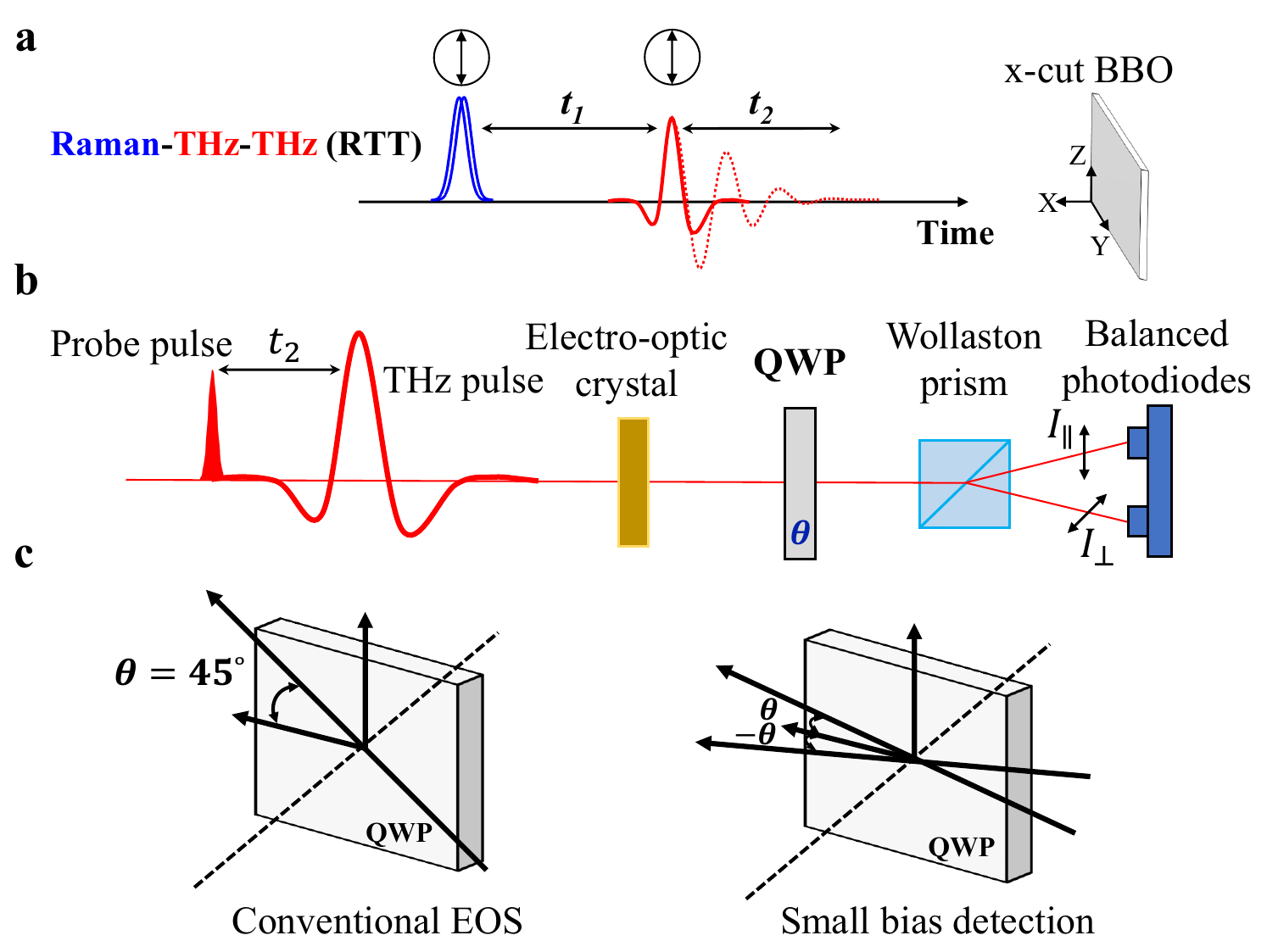}
		\caption{a) Schematic of the 2D Raman-THz spectroscopy with RTT pulse sequence. The polarization of the Raman and THz pulses is shown in the circles. b) Electro-optic sampling (EOS) detection scheme. c) Angle of the quarter-wave plate (QWP) at $45^\circ$ in conventional EOS (left panel), and two small opposite angles around zero in small bias detection scheme (right panel).}\label{Fig1}
\end{center}
\end{figure}

The experimental setup for 2D Raman-THz spectroscopy has been described in detail before.\cite{Savolainen_2013} Briefly, the output of an amplified Ti:sapphire laser system, operating at a center wavelength of 800 nm with pulse duration of 110~fs and repetition rate of 5~kHz, is divided into Raman and THz branches using a beamsplitter. The beam used for the THz branch is split further into two beams: one for THz generation and the other for THz detection. All three laser beams have parallel polarizations, aligned parallel to the crystallographic z-axis of BBO as shown in Fig.~\ref{Fig1}a. Broadband THz pulses are generated via optical rectification in a 200~$\mu$m-thick (110) GaP crystal. Before reaching the GaP crystal, the THz generation beam is modulated using a mechanical chopper operating at a frequency of one-quarter of the laser repetition rate (1.25~kHz). The generated THz pulses are first focused onto the sample and then directed towards the detection crystal (a 200~$\mu$m-thick GaP crystal) using two custom-made aluminum elliptical mirrors. The Raman pump pulse with energy of 40~$\mu$J is first modulated at half of the laser repetition rate (2.5~kHz) and subsequently focused onto the sample. The temporal delays $t_{1}$, the delay between the Raman pump and generated THz field, and $t_{2}$, the delay between THz generation and detection, are controlled by two step-scan motors. 2D Raman-THz-THz (RTT) pulse sequence is shown schematically in Fig.~\ref{Fig1}a. The emitted signal (red dotted line shown in Fig.~\ref{Fig1}a) is focused onto the detection crystal and measured using a conventional EOS detection scheme (see Fig.~\ref{Fig1}b). The entire THz section of the experimental setup, including the THz generation and detection components, is enclosed within a nitrogen-purged box to prevent water vapor absorption.

We also employed an alternative THz detection approach, known as the small bias detection scheme,\cite{Johnson_2014,Brunner_2014,Krauspe_2020} to effectively isolate the very weak third-order nonlinear signal from a strong second-order contribution caused by imperfect balancing in the 2D Raman-THz experiment on the x-cut BBO crystal. In the conventional EOS detection scheme (see Fig.~\ref{Fig1}b), the angle of the quarter-wave plate (QWP) is typically set to $45^\circ$ (see Fig.~\ref{Fig1}c-left panel), converting the linearly polarized probe beam into circular polarization which is subsequently split by the Wollaston prism into two orthogonal polarization components. In the small bias detection scheme, this angle is set to a small angle close to $0^\circ$, which effectively reduces the optical bias from the usual setting.\cite{Johnson_2014} In this configuration, a significant portion of the probe beam intensity, whose polarization is parallel to the probe beam polarization ($I_\parallel$), is directed towards one of the photodiodes. Balance detection can then be achieved by adjusting a variable attenuator located in front of that photodiode, thereby attenuating the main polarization component. The bias detection scheme has the same effect on increasing the sensitivity of the THz detection as the Brewster windows scheme that we had introduced previously.\cite{Ahmed_2014} However, as we will see, it introduces an additional tuning parameter that allows us to suppress the ${\chi}^{(2)}$-induced artifact.

\begin{figure}[t]
	\centering
	\begin{center}
		\includegraphics[width=0.48\textwidth]{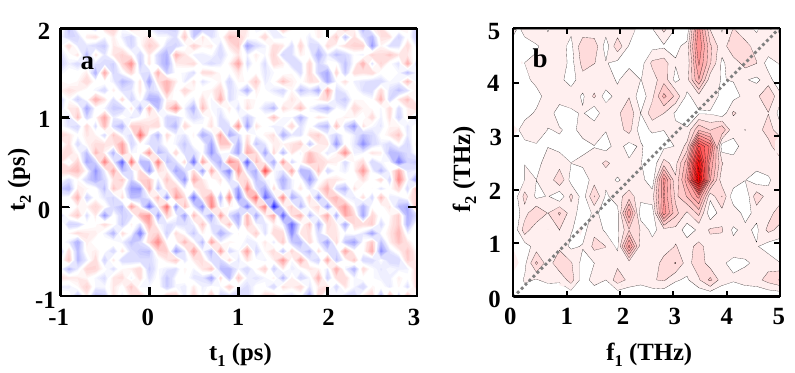}
		\caption{a) 2D time domain response measured using the RTT pulse sequence and detected by conventional EOS, for the x-cut BBO crystal. b) Absolute value of the 2D Fourier transformation for the data shown in panel (a).}\label{Fig2}
	\end{center}
\end{figure}

\section{Results and Discussion}

\subsection{2D Raman-THz spectroscopy of a x-cut BBO crystal}

Figure~\ref{Fig2}a shows the 2D time domain response measured using the RTT pulse sequence and detected by conventional EOS, for a 100~$\mu m$ thick x-cut BBO crystal. As can be seen, a long-lived oscillatory feature along both \textit{t}$_1$ and \textit{t}$_2$ axes dominates the measured signal in the time domain. Figure~\ref{Fig2}b shows the absolute value of the 2D Fourier transformation of the data in Fig.~\ref{Fig2}a. Three main peaks with frequencies of 2.2~THz, 2.85~THz, and 3.5~THz are observed along the \textit{f}$_1$ frequency axis. The observed features appear to climb up and align along the diagonal line, accompanied by a splitting of these features that follows the spectral shape of the instrument response function (IRF).  It has been previously shown that the convolution of the molecular response with the IRF strongly affects the 2D Raman-THz response in terms of spectral shape, intensity, and frequency position.\cite{Ciardi_2019} The IRF has two nodal lines, one on the diagonal and the other along the $f_{1}$ axis, suppressing diagonal peaks and those along the $f_{1}$ axis with $f_2=0$. 

In the next section, we will show that the observed features in Fig.~\ref{Fig2}b are a ${\chi}^{(2)}$-induced artifacts originating from imperfect balancing in the conventional EOS detection scheme.

\subsection{THz signal at arbitrary quarter-wave plate angle}

In the conventional EOS detection scheme, where the QWP is oriented at $\theta=45^\circ$, the intensities of the two orthogonal polarization components at the balanced photodiodes are:\cite{lee_2009} 
\begin{align}
	I_\perp= \frac{I_{0}}{2}\ (1+\sin(\phi))\\
	I_\parallel= \frac{I_{0}}{2}\ (1-\sin(\phi))
\end{align}
where $I_{0}$ is the probe beam intensity, and $\phi$ is the phase retardation experienced by the probe beam due to the THz electric field. As a result, the normalized EOS signal, which is obtained as the intensity difference between two balanced photodiodes is:\cite{lee_2009}
\begin{align}
	I_s = \frac {I_\perp - I_\parallel}{I_{0}} = \sin(\phi) 
\end{align}
However, when a QWP is oriented at an arbitrary angle $\theta$, the intensities of the two orthogonal polarization components are given by:\cite{Brunner_2014} 
\begin{align}\label{I_perpendicular}
	\begin{split}
		I_\perp (\phi,\theta) &= \frac{I_{0}}{2} \times \{1+\sin (2\theta) \sin(\phi) \\
		&\quad - \cos^2 (2\theta) \cos(\phi)\}
	\end{split}
\end{align}
\begin{align}
	\begin{split}
		I_\parallel (\phi,\theta) &= \frac{I_{0}}{2} \times \{1-\sin (2\theta) \sin(\phi) \\
		&\quad + \cos^2 (2\theta) \cos(\phi)\}.
	\end{split}
\end{align}
Here for simplicity, we neglect the small contributions from phase retardations due to the strain-induced birefringence of the detection crystal and the scattering background. To achieve balanced detection in the small bias detection scheme, the main polarization component ($I_\parallel$) needs to be attenuated by a factor $T\ll1$:
\begin{align}
	T = \frac{\sin^2(2\theta)}{1+\cos^2(2\theta)} 
\end{align}
in which case we have ($I_\perp = TI_\parallel$) when $\phi=0$. The normalized balanced difference signal then is:
\begin{align}\label{eq_Sig_theta}
	\begin{split}
		I_s (\phi,\theta) &= \frac{I_\perp(\phi,\theta) - TI_\parallel(\phi,\theta)}{I_\perp(0,\theta) + TI_\parallel(0,\theta)}\\
		&= \frac{1}{\sin^2 (2\theta)}\{(1+T)(\sin (2\theta) \sin(\phi)\\ 
		&\quad - \cos^2 (2\theta) \cos(\phi))
		 - T + 1\}
	\end{split}
\end{align}
When assuming a small THz-induced phase retardation $(\phi\ll1)$, we can Taylor expand $\sin(\phi)$ and $\cos(\phi)$ up to second order, simplifying  equation~\ref{eq_Sig_theta}:
\begin{align}\label{eq_Sig_theta_simp}
	\begin{split}
		I_s (\phi,\theta) &= \frac{1}{\sin^2 (2\theta)} \{(1+T)(\sin (2\theta) \phi \\
		&\quad - \cos^2 (2\theta) (1-\frac{\phi^2}{2}))
		- T + 1\}
	\end{split}
\end{align}  

In addition to balanced detection, we employ a double-chopping scheme in order to isolate the nonlinear response from other undesired signals arising from individual pulses, such as scattering effects, and optical rectification. The implementation of the {double-chopping scheme enables us to acquire four distinct measurements, denoted as $I_{t,r}$, where the subscripts $t$ and $r$ represent the states of the THz and Raman choppers, respectively (1 or 0):
\begin{gather*} 
	I_{1,1}= THz_{On} Raman_{On}\\
	I_{1,0}= THz_{On} Raman_{Off}\\
	I_{0,1}= THz_{Off} Raman_{On}\\
	I_{0,0}= THz_{Off} Raman_{Off}
\end{gather*}
 
In our double-chopping scheme, the Raman pump and THz generation beams are modulated at 1/2 and 1/4 of the laser repetition rate, respectively. The measured 2D Raman-THz signal is then calculated according to:
\begin{align}\label{eq_Sig_Calculation}
    S = \left( (I_{1,1} - I_{1,0}) - (I_{0,1} - I_{0,0}) \right)
\end{align}     

Together with equation~\ref{eq_Sig_theta_simp}, the four individual signals can be written as:  
\begin{align}
	I_{0,0} (0,\theta)&= \frac{1}{\sin^2 (2\theta)} \{(1+T)(-\cos^2(2\theta))- T + 1\}
\end{align}	
\begin{align}
	\begin{split} 
	  I_{1,0} (\phi_{THz},\theta)&= \frac{1}{\sin^2 (2\theta)} \{(1+T)(\sin (2\theta) \phi_{THz}\\ 
	  &\quad - \cos^2 (2\theta) (1-\frac{\phi_{THz}^2}{2}))- T + 1\}
\end{split}
\end{align}
\begin{align}
	\begin{split}
	   I_{0,1} (\phi_{Raman},\theta) &= \frac{1}{\sin^2 (2\theta)} \{(1+T)(\sin (2\theta) \phi_{Raman} \\
	   &\quad -\cos^2(2\theta) (1-\frac{\phi_{Raman}^2}{2}))- T + 1\} 
    \end{split}
\end{align}
\begin{align}
   \begin{split}
	   I_{1,1} (\phi_{THzRaman},\theta) &= \frac{1}{\sin^2 (2\theta)} \{(1+T)(\sin(2\theta) \phi_{THzRaman} \\ 
 	   &\quad -\cos^2(2\theta) (1-\frac{\phi_{THzRaman}^2}{2})) \\ 
 	   &\quad - T +1\} 
   \end{split}
\end{align}
where $\phi_{THz}$ and $\phi_{Raman}$ are phase retardations due to the THz excitation pulse transmitted through the sample, and the Raman-induced THz field, which is generated in a non-centrosymmetric sample (BBO crystal) due to the fact that its ${\chi}^{(2)}$ nonlinearity is non-zero, respectively.
The term $\phi_{THzRaman}$ represents the sum of the phase retardations induced by THz ($\phi_{THz}$) and Raman ($\phi_{Raman}$) pulses, along with the contribution from the desired third-order nonlinear signal $\phi_{3rdNL}$ (${\chi}^{(3)}$ response):   
\begin{align}\label{eq_gama}
	\phi_{THzRaman}= \phi_{THz}+\phi_{Raman}+\phi_{3rdNL} 
\end{align}
Using the above definitions for the individual signals, the measured 2D Raman-THz signal can be calculated as follows:
\begin{align}\label{eq_Sig_bias_2}
	\begin{split}
	    S &= \frac{1}{\sin^2 (2\theta)} \{(1+T)(\sin(2\theta) \phi_{3rdNL} \\
	    &\quad + \cos^2(2\theta) (\phi_{THz}\phi_{Raman}
	    + \phi_{THz}\phi_{3rdNL} \\
	    &\quad + \phi_{Raman}\phi_{3rdNL} 
	    + \frac{(\phi_{3rdNL})^2}{2}))\}  
	\end{split}    
\end{align}

\noindent Theoretically, when the angle of the QWP is set to $45^\circ$, the $\cos^2(2\theta)$-terms become zero and the signal reduces to:
\begin{align}
	S =\frac{1+T}{\sin (2\theta)} \phi_{3rdNL}
\end{align}
However, in practice, there is always some leakage due to the imperfections in the probe beam polarization and the polarizing optics. As we show in this paper, this issue becomes particularly crucial when trying to measure the weak ${\chi}^{(3)}$ signals of non-centrosymmetric materials that are ${\chi}^{(2)}$ active. When a substantial Raman-induced THz field $\phi_{Raman}$ is present, the $\phi_{THz}\phi_{Raman}$ term (called a product field) is larger than any of other terms in the equation~\ref{eq_Sig_bias_2}. Therefore, signal $S$ is dominated by the product field: 
\begin{align}
	S \propto \phi_{THz} \phi_{Raman}
\end{align}
regardless of the angle $\theta$ of the QWP, making it impossible for the conventional EOS detection scheme to isolate the weak ${\chi}^{(3)}$ signal from the strong ${\chi}^{(2)}$ artifact.

\begin{figure}[t]
	\centering
	\begin{center}
		\includegraphics[width=0.47\textwidth]{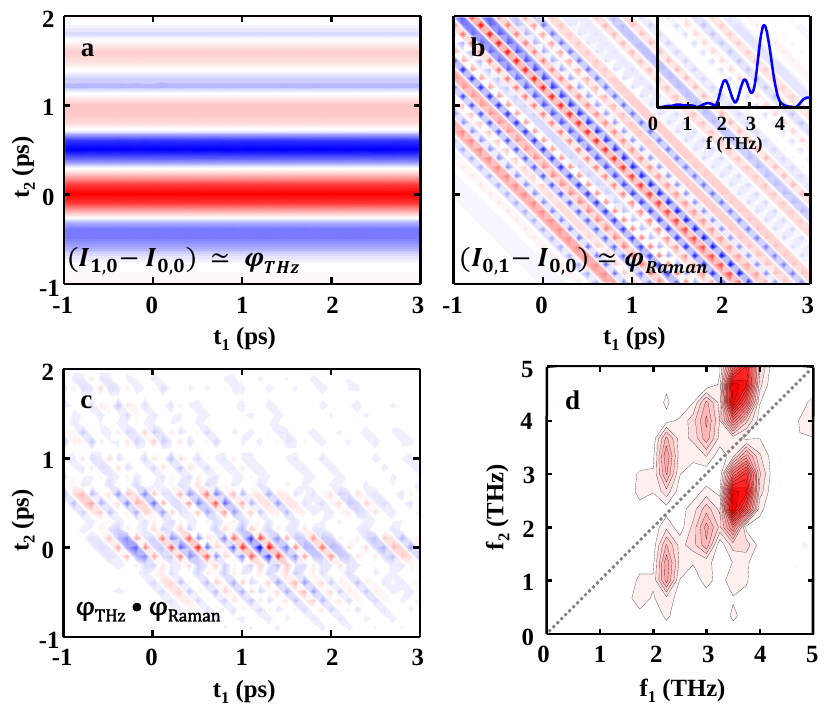}
		\caption{2D plots in the time domain for a) the THz field transmitted through the 100 $\mu m$ thick x-cut BBO crystal, denoted as $\phi_{THz}$, and b) Raman-induced THz field, labeled as $\phi_{Raman}$. Both are 1D signals, which are plotted in a 2D representation with the delay times as they are given by the experimental configuration. The inset in Fig.~\ref{Fig3}b shows the Fourier transformation of  $\phi_{Raman}$. Multiplication of the transmitted THz field through the 100~$\mu m$ thick x-cut BBO crystal and Raman-induced THz field, called product field ($\phi_{THz}\phi_{Raman}$), is shown c) in the time domain and d) in the frequency domain.}\label{Fig3}
	\end{center}
\end{figure}

Figures~\ref{Fig3}a and ~\ref{Fig3}b display in the time domain the THz field transmitted through a 100 $\mu m$ thick x-cut BBO crystal ($\phi_{THz}$), and Raman-induced THz field ($\phi_{Raman}$) respectively. Both are plotted in a 2D representation with the delay times as they are given by the experimental configuration, to help illustrate how the product field in Fig.~\ref{Fig3}c is formed. However, they are really 1D signals, in the sense that they depend on one time-variable only; in Fig.~\ref{Fig3}a on $t_2$ and in Fig.~\ref{Fig3}b on $t_1+t_2$. The inset in Fig.~\ref{Fig3}b shows the Fourier transform of the data in Fig.~\ref{Fig3}b. When the Raman pulse hits the BBO sample, it generates a Raman-induced THz field $\phi_{Raman}$ through an optical rectification process with three distinct frequency components at 2.2~THz, 2.85~THz, and 3.5~THz, which are phonon modes of the crystal.\cite{Valverde2017, Jang2020}

In order to experimentally validate the theory discussed above, we multiply the transmitted THz field through the BBO with that induced by Raman pulse, $\phi_{THz}\phi_{Raman}$, as shown in Fig.~\ref{Fig3}c. Figure~\ref{Fig3}d shows the absolute value of the 2D Fourier transformation of the data in Fig.~\ref{Fig3}c. By comparing Fig.~\ref{Fig2}b and Fig.~\ref{Fig3}d, one can clearly see that the measured signal of the BBO crystal using conventional EOS is dominated by the product term  $\phi_{THz} \phi_{Raman}$, revealing a close agreement between the two signals in both time and frequency domains. Thus, this comparison provides conclusive evidence that the measured data for BBO is not a real ${\chi}^{(3)}$ signal; rather, it is a {${\chi}^{(2)}$-induced artifact facilitated by imperfect QWP balancing. An additional difficulty in recognizing this artifact as such originates from its multiplicative nature.  Following the intensity dependence of measured signal is one of the conventional approaches to discriminating real and spurious contributions in multidimensional spectroscopies. However, it is not applicable here, as both the desired third-order nonlinear signal and the ${\chi}^{(2)}$-induced artifact are following the same intensity dependence of $I_S \propto I_{Raman} E _{THz}$.  It is worth noting that the same signal remains even as  $\theta$  is set as close as possible to $45^\circ$ due to the imperfect polarization (Fig.~\ref{Fig2}). Therefore, in 2D Raman-THz experiments involving non-centrosymmetric materials like BBO, where ${\chi}^{(2)}$ is nonzero, relying solely on the double-chopping scheme is insufficient to recover the desired third-order nonlinear signal. 

\subsection{Measurements at  $\pm\theta$ around $\theta = 0^\circ$}

In the previous section, we demonstrated both theoretically and experimentally that the measured signal in 2D Raman-THz experiments involving non-centrosymmetric samples is overwhelmingly dominated by a ${\chi}^{(2)}$-induced artifact. Here, to recover the weak third-order 2D Raman-THz response of the BBO crystal, we use the small bias detection scheme\cite{Johnson_2014,Brunner_2014,Krauspe_2020}. In this scheme, the balanced difference signal is measured at opposite optical biases $\pm\theta$ where $\theta$ is a small angle of the QWP close to $0^\circ$, as shown in Fig.~\ref{Fig1}c, right panel. 
The following procedure was used for the data acquisition. First, the $\theta$ = 0 position of
the QWP stage was determined, as explained in the Appendix, where it is also shown that the $\theta$ = 0 position needs to be accurate within $\approx0.005^\circ$. Next, for each delay time, two measurements were taken at $\theta=+2^\circ$ and  $\theta=-2^\circ$, using a motorized rotation mount (PI M-060.DG) with an angle accuracy of ±$0.001^\circ$. For each angle of the QWP, the signal was calculated according to equation~\ref{eq_Sig_Calculation}. Finally, a difference signal was obtained by subtracting these two consecutive measurements at each delay time. 
In that difference signal, the sign of $\sin (2\theta)$ coefficient flips, while the sign of $\cos^2(2\theta)$ coefficient remains unchanged (see equation~\ref{eq_Sig_bias_2}). This is not the case for two measurements taken around $45^\circ$ in the conventional EOS detection scheme. The real third-order nonlinear signal $S^{(3)}$ can thus be successfully recovered:
\begin{align}\label{deltaS}
	S^{(3)} = S(\phi,\theta)-S(\phi,-\theta) = \frac{2(1+T)}{\sin (2\theta)}\phi_{3rdNL}.
\end{align}

The result is shown in Fig.~\ref{Fig4}a as a 2D plot.  Figure~\ref{Fig4}b compares a 1D cut along \textit{t}$_{1}$ axis at \textit{t}$_{2}=0$ for the 2D difference signal (blue dotted line) with the signal that is dominated by the product field $\phi_{THz}\phi_{Raman}$  measured at positive optical bias (in red).  It is important to note that these two data sets were extracted from the same measurement, enabling a meaningful comparison. 

\begin{figure}[t]
	\centering
	\begin{center}
		\includegraphics[width=0.49\textwidth]{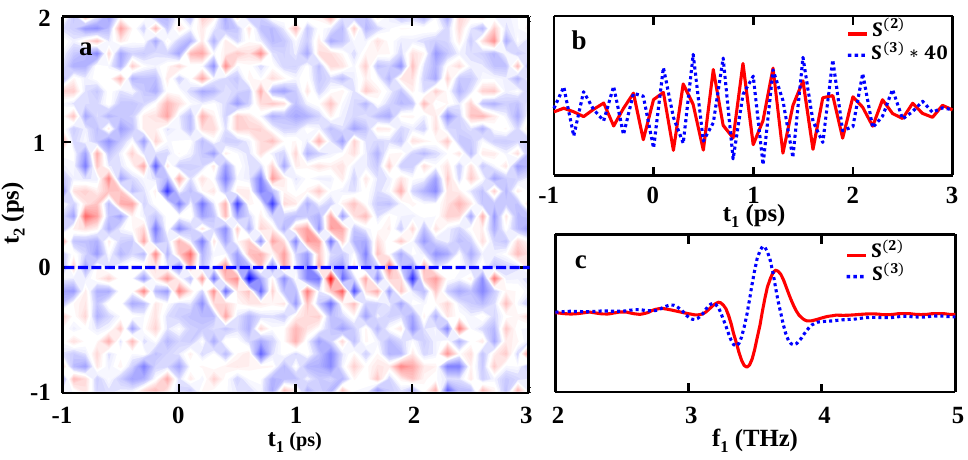}
		\caption{a) Time-domain 2D Raman-THz data (2D difference signal) for the 100~$\mu m$ thick x-cut BBO crystal obtained through the subtraction of the measured 2D signals at opposite optical biases. b) Comparison of 1D cuts along \textit{t}$_{1}$ axis for the $S^{(2)}$ artifact measured at a positive optical bias (in red), and the $S^{(3)}$ signal (in blue), which is obtained through the subtraction of the measured signals at opposite optical biases. c) Fourier transformation of the 1D time domain cuts shown in panel (b), which highlights a distinct $\pi$/2 phase shift between the $S^{(3)}$ difference signal and the $S^{(2)}$ artifact.}\label{Fig4}
	\end{center}
\end{figure}

The first observation is that the 2D difference signal ($S^{(3)}$ signal) is significantly weaker, by a factor of $\approx 40$, compared to the ${\chi}^{(2)}$-induced artifact measured at opposite optical biases (see Fig.~\ref{Fig4}b).  More importantly, the difference signal is phase-shifted compared to the ${\chi}^{(2)}$-induced artifact measured at positive optical bias. The phase difference between the two signals, determined by comparing the angles of their Fourier transforms at the peak frequency, is $\Delta\phi = 90.4 \pm 10^\circ$. To better visualize this phase shift,  the real part of the Fourier transform of the 1D time-domain cuts is shown in Fig.~\ref{Fig4}c. 

 The observed $\pi/2$ phase shift between the two time-domain signals indicates that they are generated through nonlinearities of different parity, namely ${\chi}^{(2)}$ vs ${\chi}^{(3)}$, because each successive order of nonlinear interaction introduces a $\pi/2$ phase shift due to the complex nature of the nonlinear polarization.\cite{BOYD_2008}
It is important to mention that imperfections in the subtraction of $S(\phi,\theta)$ and $S(\phi,-\theta)$ signals cannot result in a $\Delta S(\phi,\theta)$ with different phase characteristics, provided that the signals being subtracted have the same phase. As a linear operation, subtraction can only alter the amplitude and not the phase of the $\Delta S(\phi,\theta)$.  In the current method, the phases of both $\phi_{THz}$ and $\phi_{Raman}$ are established before the QWP and thus are $\theta$ independent. Thus, the phase difference between the two signals observed in Fig.~\ref{Fig4}b should be attributed to the intrinsic characteristics of a third-order nonlinear signal.

As discussed in our previous publications\cite{Ciardi_2019, Shalit_2021, Mousavi_2022}, the frequency positions of the observed features in the 2D Raman-THz spectrum are affected by the convolution of the sample response with the IRF. Thus, deconvolution is necessary to extract the real frequency positions of these features. The 2D Fourier transformation (FFT) of the time domain data (Fig.~\ref{Fig4}a), after deconvolution, is shown in Fig.~\ref{Fig5}b. The FFT was performed from \textit{t}$_{1}$ =  200~fs onwards to avoid contributions from the other pulse sequence at negative \textit{t}$_{1}$ times (THz-Raman-THz pulse sequence). The resulting 2D difference spectrum looks significantly different from the artifact 2D data shown in Fig.~\ref{Fig2}b. Specifically, it reveals a single peak that is sharper and lacks any above-diagonal features. To assign this single peak, we measured both polarized Raman and THz transmission spectra of x-cut BBO, which are shown in Fig.~\ref{Fig5}a and Fig.~\ref{Fig5}c, respectively. 

\begin{figure}[t]
	\centering
	\begin{center}
		\includegraphics[width=0.45\textwidth]{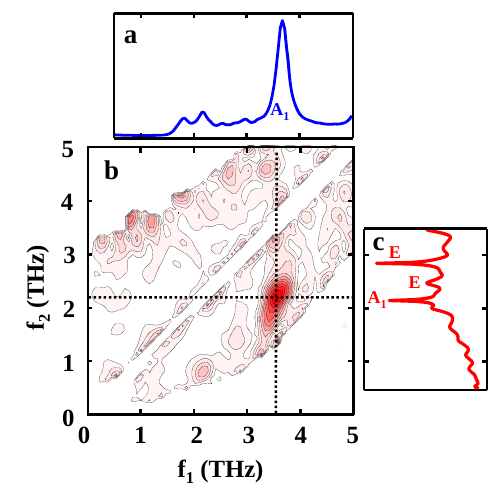}
		\caption{a) Polarized Raman spectrum of x-cut BBO with mode symmetry assignment of the main peak. b) Absolute value 2D spectrum of the $\Delta S$ signal, shown in Fig.~\ref{Fig4}a, after deconvolution. The areas where the IRF is too small, and thus deconvolution is unreliable, are zeroed out by applying a threshold of 15\% of the maximum amplitude of the IRF. c) THz transmission spectrum of x-cut BBO with mode symmetry assignment for the main peaks (sample oriented at $60^\circ$ with respect to the polarization direction of the incident THz beam, since otherwise the absorbance of the THz modes would be too strong).}\label{Fig5}
	\end{center}
\end{figure}

As can be seen from Fig.~\ref{Fig5}a, the polarized Raman spectrum of BBO reveals one strong peak at around 3.7~THz, which corresponds to the $A_1$ (TO) phonon mode of the BBO crystal and has been previously reported in the literature\cite{Ney_1998}. The THz transmission spectrum of BBO in Fig.~\ref{Fig5}c exhibits three distinct absorption features due to the resonant TO phonon modes of the BBO crystal\cite{Liu_2008}, a band at 2.15~THz which is assigned to $A_1$ phonon mode, and two bands at 2.50~THz, and 2.85~THz which correspond to $E$ phonon modes, in good agreement with the literature.\cite{Liu_2008, Liu_2009, Ney_1998} 

The frequency positions of the peak in the 2D spectrum (indicated by the black dotted lines in Fig.~\ref{Fig5}b) along the \textit{f}$_{1}$ and \textit{f}$_{2}$ frequency axes (\textit{f}$_{1} $$\approx$ 3.6~THz, \textit{f}$_{2} $$\approx$ 2.2~THz) fall very close to the positions of the two $A_1$ phonon modes observed at 3.7~THz and 2.15~THz in the polarized Raman and the THz transmission spectra, respectively. Considering that the $A_1$ mode excited by the Raman pulse can only be coupled to the modes with the same symmetry, this 2D spectrum suggests the presence of an anharmonic coupling between these two $A_1$ phonon modes within the BBO crystal.

\section{Conclusion}
In conclusion, we have presented the 2D Raman-THz response of a x-cut BBO crystal, measured using a conventional EOS detection scheme. Our comprehensive data analysis demonstrated that the observed signal is dominated by ${\chi}^{(2)}$-induced artifact due to imperfect balancing. Ignoring this effect can lead to misleading interpretations of the measured data. We furthermore showed both theoretically and experimentally that the third-order nonlinear response of the x-cut BBO crystal can be successfully recovered, effectively suppressing the strong  ${\chi}^{(2)}$-induced artifact, through the application of the small bias detection scheme\cite{Brunner_2014, Johnson_2014, Krauspe_2020}. By implementing a bias detection scheme, we measured the signal at two small opposite optical biases around the zero angle of the QWP. The resulting 2D difference signal, obtained by subtracting these two consecutive measurements, revealed a single cross-peak feature at \textit{f}$_{1} \approx$ 3.6~THz and \textit{f}$_{2} \approx$ 2.2~THz (after deconvolution). These frequencies are in good agreement with the frequency of two $A_1$ phonon modes of the BBO crystal, which are observed in the Raman and THz transmission spectra, respectively. The observed $\pi/2$ phase shift between the 2D difference signal and the artifact signal, as well as the agreement between cross-peak position and phonon mode frequencies of x-cut BBO crystal provide strong evidence that the difference signal stems from third-order nonlinear response ${\chi}^{(3)}$. Therefore, our experimental findings suggest the presence of phonon-phonon coupling within the BBO crystal. Overall, the experimental approach presented here proves to be effective in disentangling the true third-order nonlinear response from the dominant artifact arising from the ${\chi}^{(2)}$ process. This systematic approach can be applied to any other 2D THz spectroscopy experiments involving non-centrosymmetric samples, ensuring correct data interpretation. This result paves the way for further expanding the applicability of 2D Raman-THz spectroscopy, thus facilitating the exploration of anharmonic phonon couplings in nonlinear crystals.

\section{Appendix: Angular Misalignment Tolerance}

In this paper, we have demonstrated theoretically and experimentally how isolating the desired third-order nonlinear response of the sample is achievable by measuring two signals at QWP opposite orientation angles  $\theta$ and $-\theta$. In the following section, we will explore how deviations in these angles impact the feasibility of the proposed method, namely its ability to suppress strong artifact contributions to the observed signal.

Assuming that $\phi_{3rdNL}$ is smaller then both $\phi_{THz}$ and $\phi_{Raman}$, equation~\ref{eq_Sig_bias_2}  can be simplified as follows:
\begin{align}\label{eq_Sig_simplified}
	    S(\theta,\phi) &\approx \frac{(1+T)}{\sin (2\theta)} \phi_{3rdNL}  + (1+T)\cot^2(2\theta) (\phi_{THz}\phi_{Raman})
\end{align}
Let's define a mistuning parameter $\delta$  as  a deviation from $\theta =0^\circ$, such that two consecutive bias measurements occur at $\theta+ \frac{\delta}{2} $ and  - $\theta+ \frac{\delta}{2}$.  With that, equation.~\ref{deltaS} becomes: 
\begin{align}\label{S_mistuning}
      \begin{split}
	    \Delta S(\theta,\phi,\delta) &=
	     S(\theta+ \frac{\delta}{2}) - S(-\theta+ \frac{\delta}{2})\\
      &= \Bigl\{ \frac{(1+T)}{\sin (2\theta +\delta )}\phi_{3rdNL}  - \frac{(1+T)}{\sin (-2\theta+ \delta)} \phi_{3rdNL} \Bigl\}\\
      &- \Bigl\{ (1+T)\cot^2(-2\theta+\delta) (\phi_{THz}\phi_{Raman})\\
      &- (1+T)\cot^2(2\theta+ \delta ) (\phi_{THz}\phi_{Raman}) \Bigl\} 
      \end{split}
\end{align}
The terms in the first set of curly brackets in Equation.~\ref{S_mistuning} correspond to the desired $S^{(3)}$ signal, while those in the second set account for the $S^{(2)}$ artifact. By considering the experimentally obtained relative contributions of the $S^{(2)}$ and $S^{(3)}$ signals from Fig.~\ref{Fig4}b and setting $\theta = 2^\circ$, Fig.~\ref{Fig6}a illustrates on a log-log scale how these terms contribute to the total signal $\Delta S$ as a function of the mistuning parameter $\delta$. This figure reveals that while the contribution from the third-order nonlinear response (blue solid line) shows only a weak dependence on $\delta$, the artifact arising from the ${\chi}^{(2)}$-induced product field (red dashed line) steeply increases with $\delta$, overwhelming the measured signal above $\delta \approx 0.025^\circ$.

Figure~\ref{Fig6}a clearly shows that for the bias scheme to be experimentally feasible one must use a precise motor (PI  M-060.DG rotation stage has reported 0.001$^\circ$ position precision) and exactly determine the zero position of the QWP stage. To determine the  $\theta = 0^\circ$ position, the intensity on the  $I_\perp$ photodiode is measured while the QWP stage is scanned in the absence of the generated THz field ($\phi = 0$), in which case equation~\ref{I_perpendicular} reduces to :   
\begin{align}
	\begin{split}
		I_\perp (\phi = 0 ,\theta) \propto \sin^2 (2\theta) 
	\end{split}
\end{align}
Fitting the measured  $I_\perp$ intensities as a function of $\theta$ (Fig.~\ref{Fig6}b, black dots) to the expected $sin^2 (2\theta)$-dependence (red curve) is used to determine the $\theta = 0^\circ$ position. 

 \begin{figure}[t]
	\centering
	\begin{center}
		\includegraphics[width=0.33\textwidth]{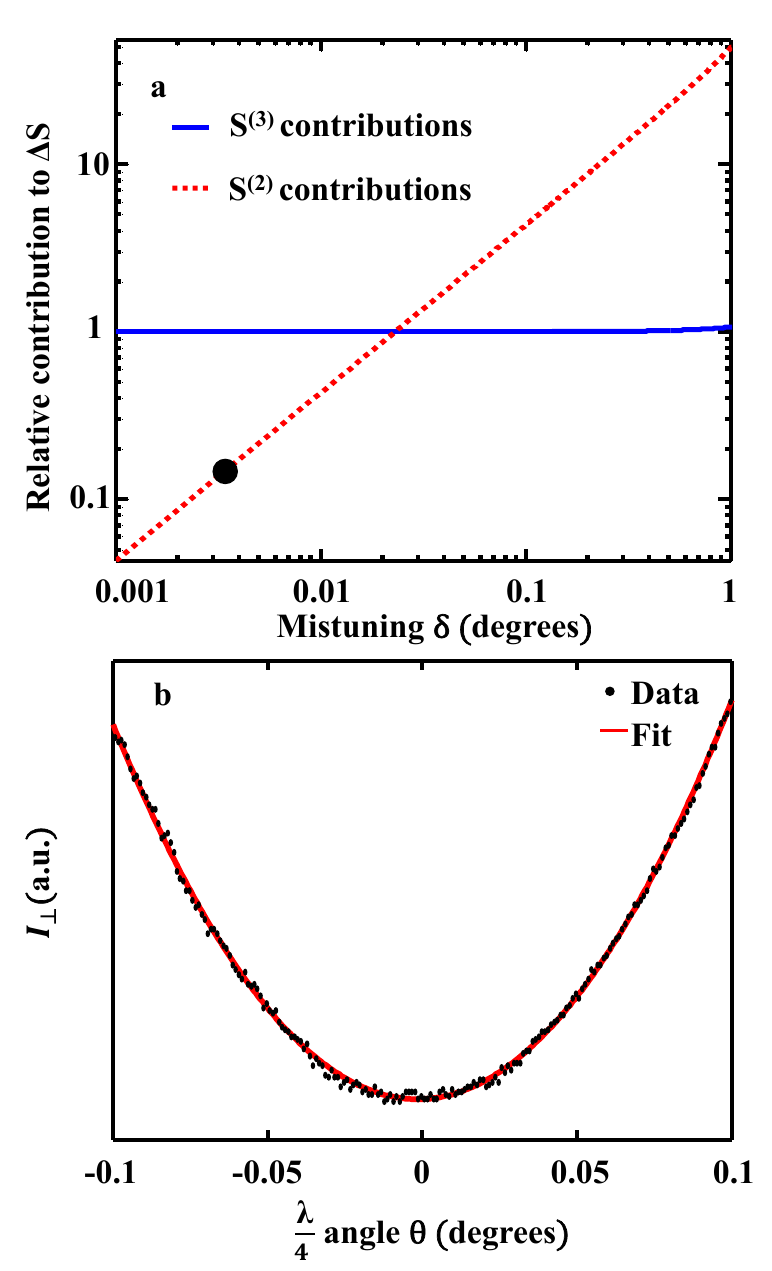}
		\caption{a) Relative contributions of ${\chi}^{(2)}$-induced artifact (red dotted line) and third-order nonlinear signal (blue solid line) to the measured signal as a function of mistining $\delta$ at  $\theta$ = $2^\circ$. The black dot shows the estimated contributions of ${\chi}^{(2)}$-artifact to the signal. b) Intensity of $I_\perp$ channel as a function of the QWP angle $\theta$ (black dots) around $\theta = 0^\circ$ and fitted $\sin^2 (2\theta)$ (red line).}\label{Fig6}
	\end{center}
\end{figure}

Finally, to evaluate the accuracy of that procedure, we make use of the phase difference of $\Delta\phi = 90.4 \pm 10^\circ$ retrieved from Fig.~\ref{Fig4}b. Together with the relative influence of the artifact on the total signal depicted in Fig.~\ref{Fig6}a, it is possible to estimate the mistuning parameter $\delta$ for the current experiment. To that end, for simplicity, we represent the artifact signal as $S^{(2)} = \sin(\omega t)$ and the real signal as $S^{(3)} = \cos(\omega t)$ (to maintain the desired $\pi/2$ phase shift). Combining both with a scaling parameter $\kappa$, $S = \kappa \sin(\omega t) + \cos(\omega t)$, a phase of $80^\circ$ (the lowest confidence interval of $\Delta\phi$) is obtained when $\kappa = 0.15$. Hence, at most 15\% of $S^{(2)}$ is mixed with $S^{(3)}$ in the final measurement. The black dot in Fig.~\ref{Fig6}a represents that value, projecting a mistuning uncertainty of $\delta \approx \pm 0.004^\circ$ for our experiment.


\vspace{0.5cm}\noindent\textbf{Acknowledgement:} The work has been supported by the Swiss National Science Foundation (SNF) through the National Center of Competence and Research (NCCR) MUST as well by the MaxWater network of the Max Planck Society.\\

%
%


\bibliography{libfile}

\end{document}